\documentclass[useAMS,usenatbib]{mn2e}

\usepackage{graphicx,amssymb,bm}
\usepackage{subfig,xfrac}
\usepackage{float, Abbrev}
\usepackage[fleqn]{amsmath}  
\usepackage{color}

\usepackage{multirow}

\newcommand{\be}{\begin{equation}}
\newcommand{\ee}{\end{equation}}
\newcommand{\ba}{\begin{eqnarray}}
\newcommand{\ea}{\end{eqnarray}}

\def\simlt{\lower.5ex\hbox{$\; \buildrel < \over \sim \;$}}

\newcommand{\fig}{\begin{figure} \begin{center}}
\newcommand{\efig}{\end{center}\end{figure} }
\newcommand{\figs}{\begin{figure*}\begin{minipage}{180mm} \begin{center}}
\newcommand{\efigs}{\end{center}\end{minipage}\end{figure*} }
\def\simgt{\lower.5ex\hbox{$\; \buildrel > \over \sim \;$}}

\title[Dark matter laboratories]{The impact of cored density profiles on the observable quantities of dwarf spheroidal galaxies}
\author[D. Harvey]
{David Harvey$^{1}$\thanks{e-mail: {\tt david.harvey@epfl.ch}}, Yves Revaz$^{1}$ , Andrew Robertson$^{2}$ and Loic  Hausammann$^1$ \\
$^{1}$Laboratoire d'Astrophysique, EPFL, Observatoire de Sauverny, 1290 Versoix, Switzerland\\
$^{2}$Institute for Computational Cosmology, Durham University, South Road, Durham DH1 3LE, UK}
\begin{document}

\date{Accepted ---. Received ---; in original form \today.}

\pagerange{\pageref{firstpage}--\pageref{lastpage}} \pubyear{2017}

\maketitle

\label{firstpage}

\begin{abstract}
We modify the chemo-dynamical code GEAR to simulate the impact of self-interacting dark matter on the observable quantities of 19 low mass dwarf galaxies with a variety star forming properties. We employ a relatively high, velocity independent cross-section of $\sigma/m=10\,\rm{cm^2/g}$ and extract, in addition to integrated quantities, the total mass density profile, the luminosity profile, the line-of-sight velocities, the chemical abundance and the star formation history. We find that despite the creation of large cores at the centre of the dark matter haloes, the impact of SIDM on the {\it observable} quantities of quenched galaxies is indiscernible, dominated mostly by the stochastic build up of the stellar matter. As such we conclude that it is impossible to make global statements on the density profile of dwarf galaxies from single or small samples.  Although based mostly on quenched galaxies, this finding supports other recent work putting into question the reliability of inferred cored density profiles that are derived from observed line-of-sight velocities.
\end{abstract}

%This finding supports other recent work putting into question the reliability of observations of cored density profiles that are derived directly from line-of-sight velocities and the small scale crisis in cosmology.


\section{Introduction}
%Dwarf spheroidal galaxies exhibit extremely low luminosities and very high mass to light ratios.
%These two key properties makes them seemingly ideal laboratories to study dark matter.
%and
%as a result they currently lie at the heart of a potential ``small-scale crisis" in cosmology. 

For the last twenty years it has been argued two major observations have caused a``small-scale crisis'' in cosmology.
%where two major observations of dwarf spheroidal galaxies are in tension with theoretical cold dark matter (CDM) only simulations. 
The first is that the central regions of observed dwarf spheroidals have flat density profiles (a core), where CDM predicts centrally-rising profiles (a cusp) \citep{dubinski1991,navarro1996,navarro1997} and the second is that there appears to be insufficient dark matter in the most massive sub-haloes \citep{klypin1999,moore1999,boylankolchin2011}. 
However, more recently it has been noted that these structures are very sensitive to baryonic physics and environmental processes, altering the total density profile changing a cusp to a core \citep{pontzen2014} and tidal stripping and ram pressure can disrupt halos bringing the expected number of observed small halos down considerably \citep{sawala2016b,wetzel2016}.  
On the other-hand, even taking into account the unknown effects of baryons, some studies of dwarf galaxies have argued that discrepancies may possibly persist \citep{schneider2017} and these inconsistencies can be attributed to an incomplete description of dark matter \citep{spergel2000,lovell2012,zavala2013,rocha2013}. As a result it remains unclear whether there exists such a ``crisis'' or whether observations have been misinterpreted \citep{verbeke2017}. %Moreover it remains unclear whether direct observations of dwarf galaxies actually have the discriminating power to unveil the nature of dark matter.

%Despite the current status of dwarf galaxies in the literature, the potential to  discriminate and probe different dark matter models is evidently clear. However, if we are to make robust conclusions we must carry out careful observationally matched studies using state-of-the-art simulations.

 In this letter we modify a suite of dwarf spheroidal galaxy simulations to simulate the impact of self-interacting dark matter (SIDM) on their observable quantities and quantify exactly how well these objects can be used to understand the nature of dark matter.
 
SIDM in recent times has been of increased interest both theoretically and observationally. Theoretically it can thermalise halos, creating cores, and also potentially reduce the amount of substructure \citep{rocha2013,peter2013,buckley2014,vogelsberger2016}, and hence ease the apparent tensions.
 Observationally, high-resolution space-based imaging of large samples of galaxy clusters has meant that now we can probe this fundamental property down to $\sigma/m\sim 1\,\rm{cm^2/g}$ \citep[e.g.][]{harvey2015,robertson2017,markevitch2004,massey2017}, however, tight constraints from low mass halos remain sparse.

\begin{table*}
\centering
\begin{tabular}{|c|c|c|c|c|c|c|c|c|c}
Model& $L_v$ & $M_\star$ & $M_{\rm 200}$ &$M_{\rm gas}$& $R_{\rm 200}$& $\sigma_{\rm LOS}$ & [Fe/H] & SF Class \\
 ID & [$10^6 L_\odot$] & [$10^6 M_\odot$] & [$10^9 M_\odot$] & [$10^6 M_\odot$] & [kpc] & [km/s] & [dex] & \\
\hline
h019 & 291/283 & 434/404 & 9.5/9.4 & 280/315 & 50.7/50.5 & 30.5/28.8 & -0.6/-0.5 & Sustained \\ 
h050 & 4.2/7.1 & 9.6/13.7 & 2.6/2.7 & 15.1/20.2 & 33.0/33.2 & 10.4/10.8 & -1.3/-1.3 & Extended \\ 
h070 & 2.0/2.0 & 5.8/5.9 & 1.8/1.8 & 0.0/1.5 & 29.2/29.3 & 10.8/10.3 & -1.5/-1.4 & Extended \\ 
h132 & 0.7/0.7 & 2.1/2.1 & 0.9/0.9 & 0.5/0.6 & 23.1/23.0 & 9.3/9.0 & -2.1/-2.0 & Quenched \\ 
h074 & 0.5/0.5 & 1.3/1.2 & 0.7/0.7 & 0.4/0.5 & 21.2/21.1 & 9.1/8.5 & -2.1/-2.2 & Quenched \\ 
h159 & 0.4/0.4 & 1.1/1.0 & 0.7/0.7 & 0.0/0.0 & 21.0/21.0 & 8.9/9.5 & -2.3/-2.3 & Quenched \\ 
h064 & 0.4/0.4 & 1.1/1.1 & 1.9/1.8 & 5.2/5.0 & 29.5/29.3 & 9.7/8.7 & -1.9/-1.9 & Quenched \\ 
h059 & 0.3/0.2 & 0.7/0.6 & 2.1/2.1 & 4.2/4.3 & 30.5/30.5 & 9.3/8.9 & -2.4/-2.2 & Quenched \\ 
h141 & 0.2/0.2 & 0.6/0.6 & 0.8/0.8 & 0.5/0.7 & 21.8/21.9 & 8.3/8.1 & -2.2/-2.4 & Quenched \\ 
h061 & 0.2/0.2 & 0.5/0.6 & 1.9/2.0 & 3.4/3.6 & 29.8/30.0 & 9.1/9.6 & -2.1/-2.1 & Quenched \\ 
h111 & 0.2/0.2 & 0.5/0.4 & 1.1/1.1 & 0.1/0.3 & 24.6/24.6 & 10.4/10.5 & -2.4/-2.4 & Quenched \\ 
h177 & 0.2/0.2 & 0.5/0.5 & 0.5/0.5 & 0.0/0.0 & 19.4/19.4 & 7.7/7.9 & -2.6/-2.3 & Quenched \\ 
h091 & 0.2/0.2 & 0.4/0.4 & 1.4/1.3 & 0.8/0.7 & 26.5/26.3 & 10.1/8.2 & -1.9/-2.5 & Quenched \\ 
h106 & 0.2/0.1 & 0.4/0.3 & 1.1/1.1 & 0.1/0.1 & 24.6/24.5 & 9.9/8.3 & -2.8/-2.3 & Quenched \\ 
h122 & 0.1/0.1 & 0.4/0.3 & 1.0/1.0 & 0.0/0.0 & 23.7/23.7 & 9.1/9.2 & -2.4/-2.5 & Quenched \\ 
h104 & 0.1/0.1 & 0.3/0.4 & 0.9/0.9 & 0.3/0.1 & 23.3/23.0 & 9.0/8.5 & -2.3/-2.4 & Quenched \\ 
h123 & 0.1/0.2 & 0.3/0.4 & 0.9/0.9 & 0.1/0.0 & 23.2/23.1 & 7.6/7.6 & -2.2/-2.3 & Quenched \\ 
h180 & 0.1/0.1 & 0.3/0.3 & 0.3/0.3 & 0.0/0.0 & 15.9/15.7 & 6.9/6.9 & -2.4/-2.3 & Quenched \\ 
h168 & 0.1/0.1 & 0.3/0.3 & 0.6/0.6 & 0.0/0.1 & 19.7/19.6 & 8.3/9.0 & -2.6/-2.5 & Quenched \\ 
\hline
\end{tabular}
\caption{A direct comparison between 19 simulated dwarf galaxies with CDM and the corresponding halo with SIDM (i.e. CDM / SIDM). From left to right the columns give the halo ID, the total $V$-band luminosity within $R_{200}$ , the total stellar mass within $R_{200}$, the total mass within $R_{200}$, the total gas mass within $R_{200}$, $R_{200}$  (the radius at which the mean enclosed density is 200 times the critical density), the mean line of sight velocity and the metallicity. The final column gives the classification of star formation whether it is sustained, extended or quenched.
\label{tab:data}}
\end{table*}

Given the nature of Dwarf Spheroidals and the peak in interest in SIDM, the number of studies looking at simulations of SIDM at the dwarf galaxy scale has naturally grown. \cite{vogelsberger2014} originally studied two dwarf galaxies with both a constant and velocity dependent cross-section. Studying relatively large dwarf galaxies ($M_{\rm halo}\sim10^{10}M_\odot$, $M_\star\sim10^8M_\odot$) they found that self-interactions had minimal impact on the global properties of the dwarf galaxies. However they did find that the metallicity in the centre rose by 15\% and the stellar distribution traced that of the cored dark matter. Following this, the Feedback In Realistic Environments (FIRE) simulations simulated four dwarf galaxies including SIDM \citep{robles2017}. This study looked at smaller mass dwarf galaxies finding that even the smallest halos ($<3\times10^{6}M_\odot$) formed cores, which were relatively unaffected by baryonic physics.
Although seminal to SIDM work, these studies lacked two important characteristics. The first is the small sample sizes meant they could not probe the diversity of profiles within a range of dwarf galaxies from a variety of formation histories. 
%In such stochastic, highly non-linear systems, it is important to sample many galaxies from multiple formation histories. 
%In fact SIDM has been proposed as a solution to the high diversity in the light curves of spiral %galaxies \citep{SIDM_diversity}. 
Secondly these studies did not calculate the impact on the direct observables, including the line-of-sight (LOS) velocities of stars within the halos and verify that the metallicity abundances were consistent with observations.

In this letter we therefore look to extend this work, simulating a larger suite of dwarf galaxies using the chemo-dynamical code \texttt{GEAR} \citep{revaz2012,revaz2016}. 
Recently, \citet{revaz2018} demonstrated that cosmological simulations with \texttt{GEAR} reproduced the details of a wide range of observable quantities of dwarf galaxies, including LOS velocity dispersion profiles, half-light radii, star formation histories, metallicity distributions, [Mg/Fe] abundance ratios, metallicity gradients
and kinetically distinct stellar populations.
It is therefore now possible to fully test the impact of more exotic models of dark matter. In particular we will study for the first time how SIDM
influences these observables.

\section{Simulations}

\texttt{GEAR} \citep{revaz2012,revaz2016} is a fully parallel chemo-dynamical Tree/SPH code based
on \texttt{Gadget-2} \citep{springel2005}.
It includes gas cooling, hydrogen self-shielding, star formation, chemical evolution, and Type Ia and II supernova yields
and thermal blastwave-like feedback. 
\texttt{GEAR} includes recent and essential SPH improvements like the pressure-entropy formulation
\citep{hopkins2013} and operates with individual and adaptive time
steps as described in \citet{durier2012}. 

For the purpose of this study, we extend \texttt{GEAR} to include SIDM. Following \citet{robertson2017}, we implement isotropic and velocity-independent scattering of dark matter particles, using a relatively large cross-section of $\sigma/m = 10\,\rm{cm^2/g}$ to maximise any effect. 
%Although this module has been tested and validated in a comoving cosmological simulations \cite{SIDM_bullet}, 
%in order to avoid any  we carry out a brief test whereby we extract a halo from the cosmological box and evolve it in non-comoving coordinates. We match this to the same halo in comoving coordinates and find that they have the same evolved properties. 
\figs
\includegraphics[width=\textwidth]{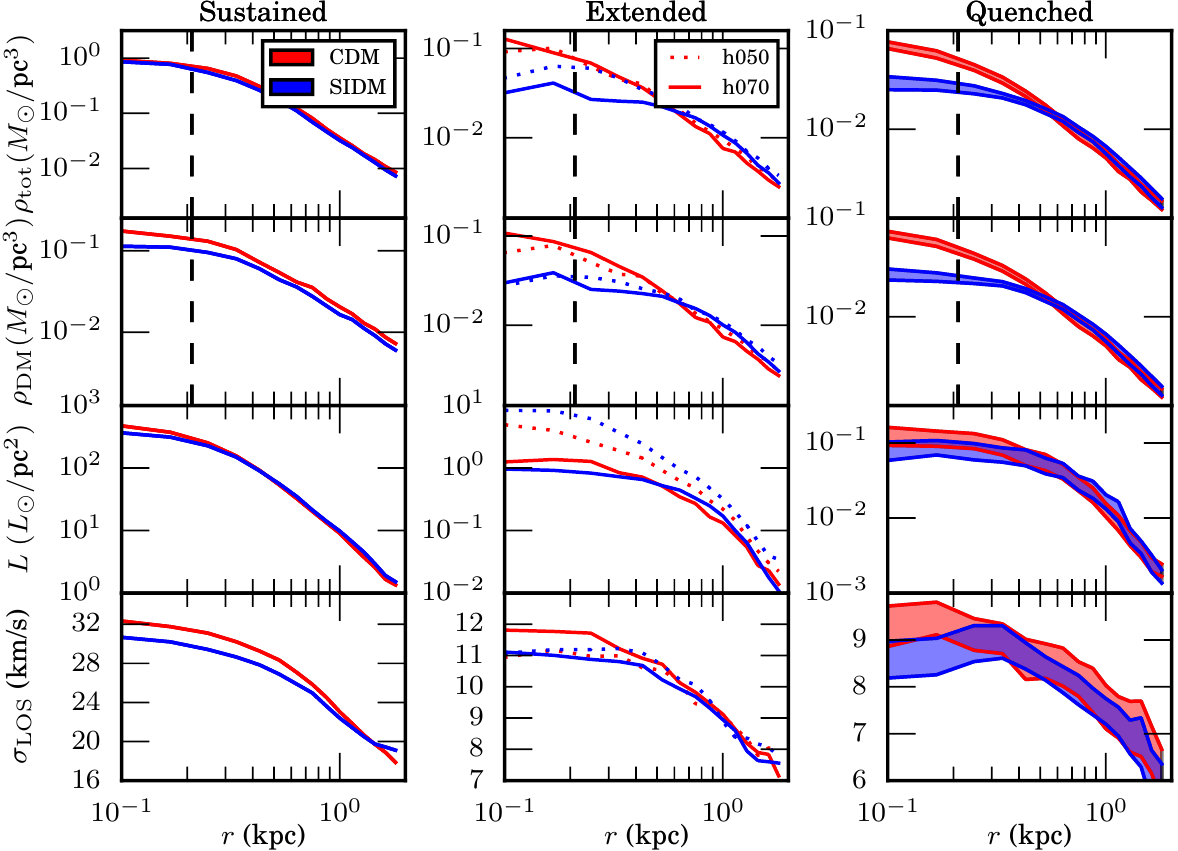}
\caption{\label{fig:ResultsA} 
Four radial dependent properties of the three samples of dwarf galaxies at $z=0$ with the CDM model in red and the SIDM in blue. The left hand column gives the sustained star forming, the middle column the two extended star forming. The final column gives the shaded region within which 68\% of the 16 quenched galaxies lie. 
The vertical dashed line corresponds to 2.8  times the softening length of the dark matter particles. From top the bottom the rows give 1.) the total mass density profile; 2.) the dark matter density profile; 3.) the lumonisity surface density and 4.) The LOS velocity profile of the stars.}
%The shaded grey region shows the softening length for the dark matter particles. The softening length for the gas and stars is $\sim20$pc and hence only show the profiles above this value. 
\efigs

\figs
\includegraphics[width=\textwidth]{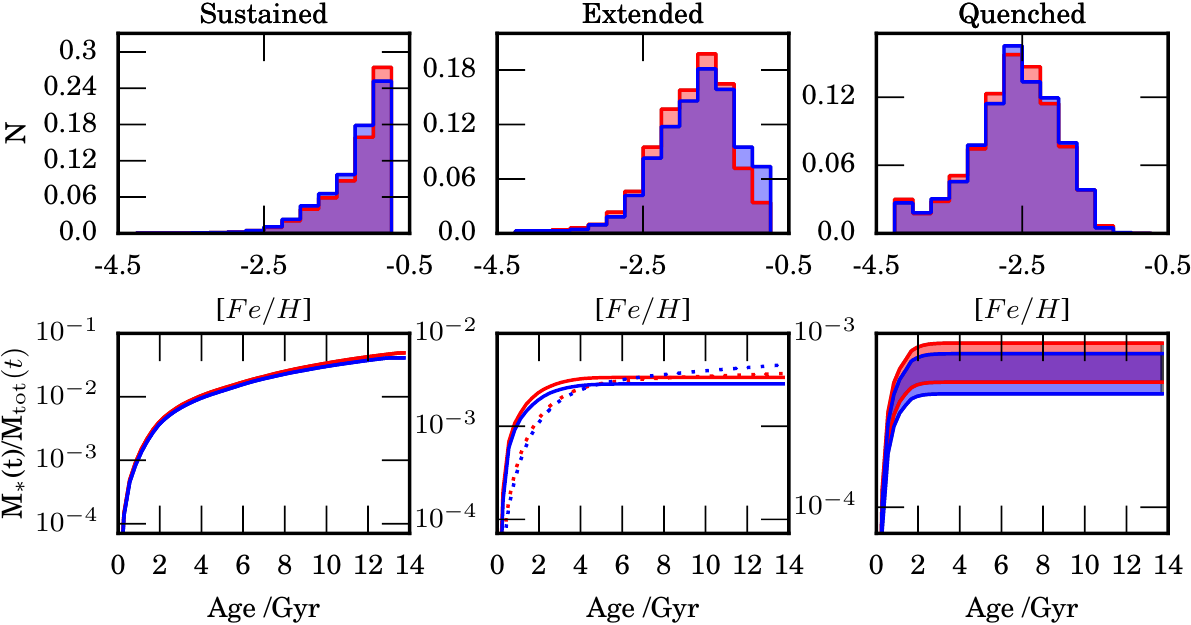}
\caption{\label{fig:ResultsB} Same columns as Figure \ref{fig:ResultsA}. The first row gives the metallicity density function and the second gives the integrated star formation history normalised to the total mass of the halo.}
\efigs

To check the impact of SIDM on dwarf galaxies' properties, we selected 19 halos from
\cite{revaz2018} and re-simulate them with our SIDM prescription.
The selection is performed to cover the three different types of star formation histories defined in 
\cite{revaz2018}, namely:  sustained, extended and quenched. 
%In practice, due to our high mass resolution ($m_\star=1024\,\rm{M_\odot}/h$) \textcolor{red}{`Due to our high mass resolution'. Are we saying that if we changed the resolution, the final stellar mass would substantially change?}\textcolor{blue}{I assume Yves just means that the mass resolution for the larger dwarfs means we have longer CPU time, but i can see how this could be misleading... is this correct Yves?}, 
Sustained  star formation histories produce bright dwarfs ($L_{\rm{V}}> 10^8\,\rm{L_\odot}$) which due to our high mass resolution ($m_\star=1024\,\rm{M_\odot}/h$) require substantial CPU resources\footnote{For those bright galaxies that continuously form stars, owing to the large sound speed of the feedback-heated dense gas, the time step may drop below $1000\,\rm{years}$.}. Conversely, quenched galaxies are faint ($L_{\rm{V}}< 10^6\,\rm{L_\odot}$) and much quicker to simulate. We therefore simulate sixteen quenched galaxies, one sustained one (\texttt{h019})
and two extended ones (\texttt{h050} and \texttt{h070}),
%Given the sensitivity of the faintest dwarf galaxies to ``external'' perturbations such as stellar feedback, UV heating or SIDM in this case, our selection, where faint galaxies are over represented, is justified.
% I DON'T THINK THE SENTENCE ABOVE ADDS ANYTHING. AND I DON'T UNDERSTAND WHAT WOULD BE `EXTERNAL' ABOUT SIDM.
in an identical manner (except the inclusion of dark matter scattering) to their counterparts in \citep{revaz2018}. % They have been run from $z=70$ down to $z=0$ using a zoom-in technique.

\section{Results}
%Here we present the results from our dwarf galaxy zoom simulations. 
Table \ref{tab:data} gives an overview of the results. For each column we show the CDM value followed by the SIDM value for the same halo. Throughout we split our sample into three star forming classes: sustained, extended and quenched as show in the final column of this table.

In Figure \ref{fig:ResultsA} we show the radial dependence of four different properties for the three samples of dwarf galaxies at $z=0$. The left hand column gives the sustained star forming galaxy, the middle column gives the two extended star forming galaxies and the final column gives the distribution of the 16 quenched galaxies, giving the region in which 68\% of these galaxies lie. %Given that the galaxies in both distributions are identical this gives an idea on the total effect of SIDM on the observable quantities of the dwarf spheroidal galaxies.
% THIS SENTENCE WOULD HAVE TO BE IMPROVED. THE GALAXIES ARE NOT IDENTICAL, THEY JUST HAD IDENTICAL INITIAL CONDITIONS
We also delineate the 2.8 times Plummer-equivalent gravitational softening length for the dark matter particles in the simulation, value beyond which the gravitational forces are Newtonian.
%We do not show this for the gas or star particles since this is $\sim20$pc and hence we show only the radial dependence above this value.

The top row shows the total matter density profile. In all but the sustained star forming galaxy we find that SIDM leads to cored density profiles, while CDM has larger central densities, also shown in the dark matter only profiles. Interestingly the core is unobservable in the total matter profile of \texttt{h019} suggesting that the additional stellar component dominates in these galaxies and is also cored. 

Investigating \texttt{h019} further, we find that the mass within 300pc doubles in the final 2~Gyrs. This effect could be due to core collapse from a very high cross-section \citep{CoreCollapse}. However, \texttt{h019} is complicated, experiencing a major merger and as such particularly difficult to interpret.
%, which is anti-intuitive. However although clearly dominating, it is difficult to conclude if the stellar distribution really is cored since it is very close to the softening length of the simulation.

%however there does appear to be a small core \textcolor{red}{do we know yet if this is numerical or physical?}\textcolor{green}{With the updated Fig. 1, the problem should have disapeared.}. 
%SIDM on the other hand exhibits a significantly larger core.
Despite the dramatic difference between CDM and SIDM, we find that the projected luminosity profiles do not differ noticeably as shown in the third row. Interestingly, we find that SIDM in \texttt{h050} has the opposite effect to what is expected. The SIDM increases the luminosity of this galaxy despite having a cored dark matter density, demonstrating the stochasticity during the formation of stars and how their dynamics are only weakly coupled to the dark matter. Finally, the quenched galaxies exhibit overlapping distributions, meaning that it would be impossible to differentiate between SIDM and CDM in these small, low mass haloes.

The final row shows the LOS velocity profiles. We find that the sustained star forming galaxy is slightly lower for the SIDM than the CDM. Although the total mass is the same, this difference might be the result of late time formation of stars within a cored halo resulting in this lower velocity. The middle column presents a somewhat unclear conclusion. Although \texttt{h070} reacts to the drop in total density by exhibiting a lower velocity profile, halo \texttt{h050} clearly has an indistinguishable velocity profile, despite also having reduced total and dark matter densities. We postulate that although these galaxies can have very different density profiles, stochastic build up of the stellar matter at early times in the central regions can cause CDM to mimic the observables of cored profiles in SIDM. The final column shows overlapping distributions, suggesting that it is common for dwarfs to have similar velocity profiles despite harbouring a factor of $\sim5$ difference in total density. 

Following these tests we carry out two further studies. The top row of Figure \ref{fig:ResultsB} shows the impact of SIDM on the metallicity distribution of stars in the galaxies. To derive these distributions we sum each metallicity density distributions at $z=0$ and took the mean. The bottom row gives integrated star formation rate normalised to the total halo mass as a function of cosmic time. We again give the region in which 68\% the galaxies lie. We find star forming cases there is no clear differentiation between the two models and that the creation of a core in the total mass profile does not alter the chemical evolution of the dwarf galaxies {\it on the whole}. The two halos that experience extended star formation have a slightly higher mean metallicity with SIDM, however the difference is only small, and so potentially unreliable with only two halos.

\subsection{Time evolution of the line-of-sight velocities\label{sec:time_vel}}
It is striking from Fig.~\ref{fig:ResultsA} that although the total mass within the central region changes significantly with SIDM, we find little change in the stellar LOS velocities. We postulate this is because the observable quantities of the stars are defined by the potential in which they are formed, when the halo harbours a cusp. To examine this further we study the time evolution of the mass within $300\,\rm{pc}$ of three dwarf galaxies and the stellar LOS velocity at the same radii. Figure \ref{fig:timeEvol} shows the results. Each panel presents the results from a single dwarf galaxy with the CDM (SIDM) results in red (blue). The shaded region shows the evolution of the LOS velocities over cosmic time with the $1\sigma$ error and the solid lines shows the total mass within a 3D radius of $300\,\rm{pc}$. We clearly see the core developing, with the SIDM mass departing from the CDM mass over time. However, the LOS velocity remains unchanged from its initial values, only mildly evolving with time. 
We conclude that the impact on the LOS velocity due to a redistribution of the central dark matter 
is imperceptible. This conclusion is confirmed by a Jeans prediction of the velocity dispersion of our simulated dwarfs.
% I don't like this sentence. At z=0 dwarfs are at equilibrium and the vel. disp. must reflect this
% dynamical equilibrum... and  not some perturbations that occured long time ago.
%We conclude from this that the line-of-sight velocity does not evolve in the same way as the mass distribution. 
%As such it is more likely that the line-of-sight velocities are defined by the stochastic build up of the galaxy %and not the gradual change in mass density. 

\fig
\includegraphics[width=0.45\textwidth]{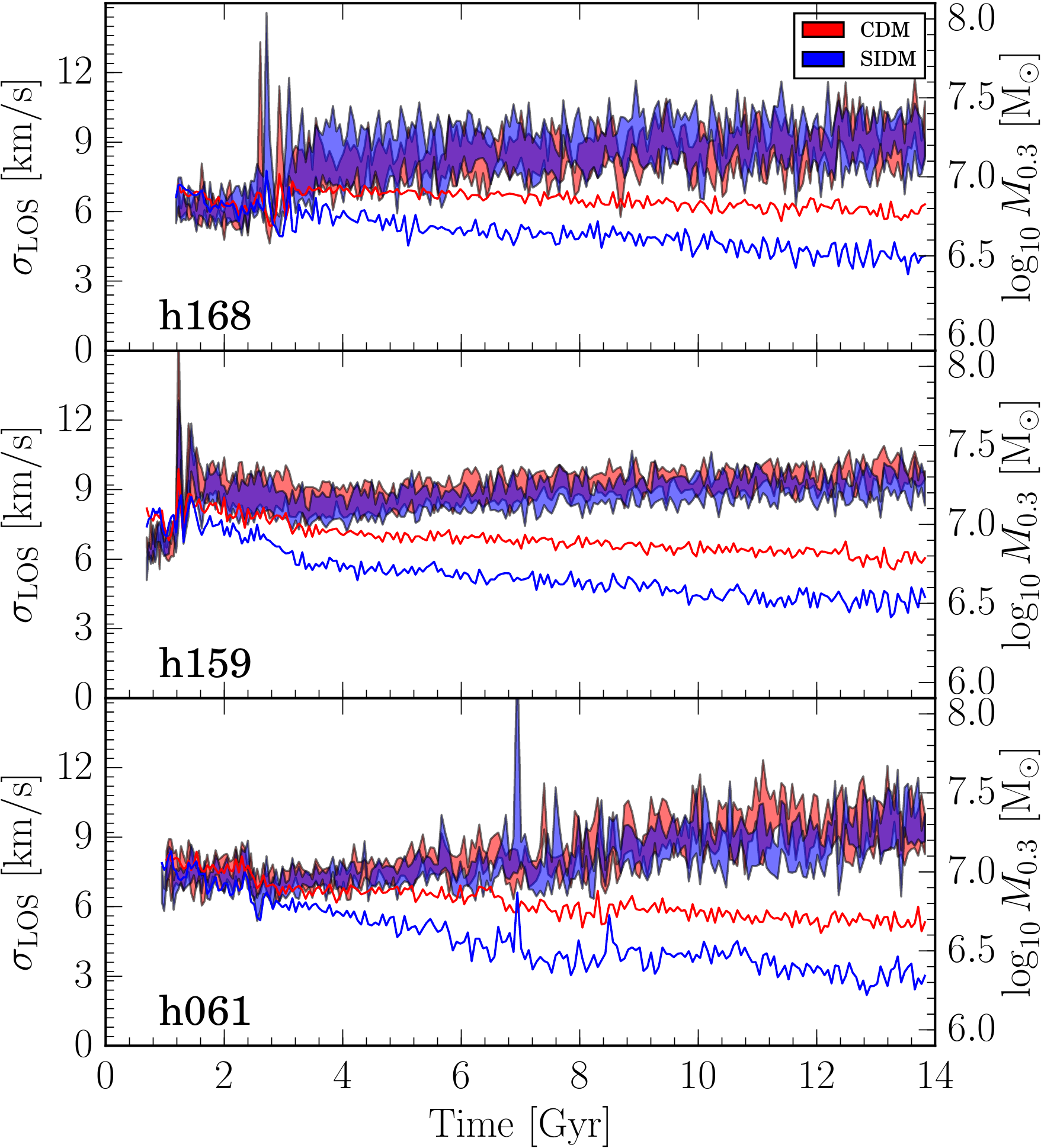}
\caption{\label{fig:timeEvol} The time evolution of the LOS velocity of stars at $500\,\rm{pc}$ from the centre of the halo and its associated $1\sigma$ error for three  dwarf galaxies. The solid line is the time evolution of the integrated mass inside the same radius. The red are the dwarf galaxies with CDM and the blue is the SIDM.}
\efig

\section{Discussion \& Conclusions}
We simulate a suite of 19 dwarf spheroidal galaxies using a modified version of the chemo-dynamical code GEAR to include self-interacting dark matter. We simulate a velocity independent cross-section of $\sigma/m=10\,\rm{cm^2/g}$ with a three regimes of star formation and extract four observable quantities: the luminosity and the line-of-sight velocity profile, the metallicity abundance and the integrated star formation history. In general we find the change in dark matter model changes the individual evolution of each galaxy, with each dwarf spheroidal forming a dark matter core in the central region.  However, despite this the distribution of the 16 quenched galaxies does not and that there is no discernible difference in observable quantities. We postulate that this due to the stochastic nature of early star formation within the galaxy and are henceforth defined by its initial potential and are insensitive to the redistribution of matter caused by SIDM. Although the limited number of extend and sustained star forming dwarf galaxies prevent us from extrapolating this conclusion to these galaxy types we do find large variances in the formation history of these galaxies and therefore raise caution to studies making global statements about dwarf galaxies based on observations of single or small samples.
Although based on our sample of quenched dwarf spheroidals, this finding is consistent with other studies of dwarf galaxies that puts in to questions the reliability of the interpretations from observations that dwarf galaxies harbour cored density profiles \citep{strigari2017} and that a small scale crisis exists in cosmology.

\section*{Acknowledgments}

This research is supported by the Swiss National Science Foundation (SNSF). D. Harvey also acknowledges support by the Merac foundation. AR is supported by a European Research Council Starting Grant (ERC-StG-716532-PUNCA).

%\bibliographystyle{mn2e}
%\bibliography{bibliography}
%\bsp
%\label{lastpage}

\bibliographystyle{aa}
\bibliography{bibliography}
\nocite{*}

\end{document}